\documentclass[runningheads]{llncs}
\usepackage{graphicx, verbatim, amssymb, amsmath, braket}
\usepackage[english]{babel}

\newcommand{\be}{\begin{equation}}
\newcommand{\ee}{\end{equation}}

\newcommand{\h}{\hat}

\newcommand{\ex}{\text{ex}}
\newcommand{\mcO}{\mathcal{O}}
\newcommand{\su}{\mathfrak{su}}

\newcommand{\Tr}{\text{Tr}}
\newcommand{\mcH}{\mathcal{H}}

\newcommand{\mbC}{\mathbb{C}}
\newcommand{\mbI}{\mathbb{I}}

\begin{document}
\title{Generalized Gibbs Ensembles in Discrete Quantum Gravity\thanks{This is an invited contribution to the conference proceedings of GSI 2019, to be published in Springer's LNCS series: Geometric Science of Information 2019, ed. F. Nielsen and F. Barbaresco.}
}
%
%
\author{Goffredo Chirco\inst{1}\orcidID{0000-0001-9538-4956} \and
Isha Kotecha\inst{1,2}\orcidID{0000-0002-3515-4355}}
%
%
\institute{Max Planck Institute for Gravitational Physics (Albert Einstein Institute), \\ Am M\"uhlenberg 1, 14476, Potsdam-Golm, Germany \and Institute for Physics, Humboldt-Universit\"{a}t zu Berlin, \\Newtonstra{\ss}e 15, 12489 Berlin, Germany \\
\email{goffredo.chirco@aei.mpg.de}\\
\email{isha.kotecha@aei.mpg.de}}
\maketitle              
\begin{abstract}
Maximum entropy principle and Souriau's symplectic generalization of Gibbs states have provided crucial insights leading to extensions of standard equilibrium statistical mechanics and thermodynamics. In this brief contribution, we show how such extensions are instrumental in the setting of discrete quantum gravity, towards providing a covariant statistical framework for the emergence of continuum spacetime. We discuss the significant role played by information-theoretic characterizations of equilibrium. 
We present the Gibbs state description of the geometry of a tetrahedron and its quantization, thereby providing a statistical description of the characterizing quanta of space in quantum gravity. 
We use field coherent states for a generalized Gibbs state to write an effective statistical field theory that perturbatively generates 2-complexes, which are discrete spacetime histories in several quantum gravity approaches.

\end{abstract}

\section{Discrete Quantum Spacetime}

From the existence of singularities in classical gravitational theory to the discovery of horizon entropies in semiclassical settings, many studies have hinted at a discrete quantum microstructure of spacetime. Precisely what these quanta of spacetime are, and how they give rise to a continuum  
gravitational field is the holy grail of non-perturbative discrete quantum gravity. It is a complex open issue, being tackled from various sides. Despite many conceptual and technical differences between the different formalisms, they admit an interesting commonality: modelling of spacetime quanta as geometric polyhedra. 

In particular, tetrahedra are the candidates of choice in 4d models for quantum excitations of geometry in several approaches, such as loop quantum gravity, spin foams, group field theory, dynamical triangulations and simplicial gravity. Collective dynamics of such degrees of freedom is then expected to give rise to an emergent spacetime. Statistical mechanical and field theoretic techniques are thus crucial from the point of view of an emergent spacetime, not only for providing tools to extract effective spacetime `macroscopic' physics from the quantum gravitational `microscopics', but also as probes to investigate non-perturbative features. A natural way for such explorations is to consider quantum spacetime as a many-body system \cite{Oriti:2017twl}, which is complementary with the view of classical continuum spacetime as an effective thermodynamic system.
%



The procedure of maximizing information entropy subject to a given set of constraints as presented by Jaynes \cite{Jaynes:1957zza,Jaynes:1957zz} is uniquely positioned to be utilised in background independent systems for defining an equilibrium Gibbs state. The primary reasons for this are that this method does not rely on the existence of any 1-parameter automorphism of the system (such as physical time evolution), unlike the customary Kubo-Martin-Schwinger condition of non-relativistic statistical mechanics. It also allows for considering observables other than energy, such as geometric volume, which may not necessarily be naturally understood as symmetry generators. These two technical features of not requiring a 1-parameter group of symmetries a priori, and an inclusion of other observables of interest, makes this procedure particularly valuable in background independent quantum gravity settings. Moreover, in an almost unassuming way, it points toward a fundamental status of information entropy in quantum gravity, which has been a recurring theme across various avenues in modern theoretical physics.



\section{Generalized Gibbs States} \label{maxent}



A macrostate of a system with many underlying degrees of freedom is given in terms of a finite number of observable averages. Jaynes \cite{Jaynes:1957zza,Jaynes:1957zz} argued that the least biased statistical distribution over the microscopics of the system, compatible with our limited knowledge of its macroscopics in terms of these averages, is that which maximizes the information entropy.
 The resultant distribution is Gibbs, which faithfully encodes our partial knowledge of the system. Maximizing the uncertainty in this manner ensures that we are using exactly only the information that we have access to, not less or more. 


Let $\{\mcO_a\}_{a=1,2,...}$ be a finite set of smooth, real-valued functions on a finite-dimensional symplectic phase space $\Gamma_\ex$. It is the unconstrained, extended state space with respect to all constraints. Let $\rho$ be a statistical density (real-valued and positive function, normalised with respect to Liouville measure) on $\Gamma_\ex$. The functions $\mcO_a$ are such that their statistical averages in $\rho$ are well-defined and constant, that is
\be \label{const1}
\langle \mcO_a \rangle_\rho \equiv \int_{\Gamma_\ex} d\lambda \; \mcO_a \,\rho \;\;=\;\; U_a \;.
\ee
Shannon entropy of $\rho$ is,
\be \label{ent}
S[\rho] = -\langle \ln \rho \rangle_\rho 
\ee
and its normalization is $ \langle 1 \rangle_\rho = 1$. Now consider maximization of $S[\rho]$ under the constraints of state normalization and equations \eqref{const1} \cite{Jaynes:1957zza}. This optimization problem can be phrased in the language of Lagrange multipliers and imposing stationarity of an auxiliary functional,
\be \label{auxfn}
L[\rho,\beta_a,\kappa] = S[\rho] - \sum_{a} \beta_a (\langle \mcO_a \rangle_{\rho} - U_a) - \kappa(\langle 1 \rangle_\rho - 1)
\ee 
with multipliers $\beta_a, \kappa \in \mathbb{R}$. Stationarity with respect to variations in $\rho$ then results in a generalized Gibbs state of the form,
\be \label{genstate1}
 \rho_{\{\beta_a\}} = \frac{1}{Z_{\{\beta_a\}}} e^{-\sum_a \beta_a \mcO_a} \ee 
 with the partition function,
 \be Z_{\{\beta_a\}} \equiv \int_{\Gamma_\ex} d\lambda \; e^{-\sum_a \beta_a \mcO_a}  = e^{1+\kappa} \ee 
where $\{\beta_a\}$ and $\mcO_a$ are such that the above integral converges.

The above procedure can be carried out analogously for finite quantum systems \cite{Jaynes:1957zz}, given that the operators under consideration have well-defined trace averages on a kinematic, unconstrained Hilbert space. Here statistical states are density operators (self-adjoint, positive and trace-class), and the ensemble averages for (self-adjoint) operators $\h{\mcO}$ are,
 \be \langle \h{\mcO_a} \rangle_\rho \equiv \Tr(\h{\rho} \, \h{\mcO}_a) = U_a \;. \ee
Then Jaynes' method gives a generalized Gibbs density operator,
 \be  \h{\rho}_{\{\beta_a\}} = \frac{1}{Z_{\{\beta_a\}}} e^{-\sum_a \beta_a \h{\mcO}_a} \;.\ee
Averages $U_a$ are generalized energies, $\beta \equiv \{\beta_a\}$ is a generalized vector-valued (inverse) temperature, and $dQ_a \equiv dU_a - \langle d\mcO_a \rangle$ are generalized heat differentials.

This information-theoretic manner of defining equilibrium statistical mechanics is to elevate the status of entropy as being more fundamental than energy. This perspective can prove instrumental in background independent settings \cite{Kotecha:2018gof}. 
As long as the system is equipped with a well-defined state space and an observable algebra, and is described macroscopically with a few observables $\{\mcO_a\}$ in terms of its averages $\{U_a\}$, the maximum entropy principle can be applied to characterize a notion of generalized statistical equilibrium.




\section{Statistically Constrained Tetrahedra}


 Jaynes' characterization of equilibrium also allows for a natural group-theoretic generalization of thermodynamics, whenever the constraint is associated to some (dynamical) symmetry of the system. In this case, the momentum map associated to the Hamiltonian action of the symmetry group on the covariant (extended) phase space of the system plays the role of a generalized energy function, comprising the full set of conserved quantities. Moreover, its convexity properties allow for a generalization of standard equilibrium thermodynamics \cite{souriau}. \ 

This approach is useful also in the simplicial geometric context of non-perturbative quantum gravity \cite{Oriti:2006se,qg}. We will use the generalized Gibbs states to define along these lines a statistical characterization of tetrahedral geometry in terms of its closure, starting from the extended phase space of a single open tetrahedron. The closure constraint is what allows to interpret geometrically a set of 3d vectors as the normal vectors to the faces of a polyhedron, and thus to fully capture its intrinsic geometry in terms of them. Subsequently, we will consider a system of many closed tetrahedra (or polyhedra in general) and demonstrate its relation to the group field theory approach to quantum gravity.




\subsection{Classical closure fluctuations}

The symplectic space $\Gamma_{\{A_I\}} = \{(X_I) \in \su(2)^{*4} \cong \mathbb{R}^{3\times 4} \;|\;  ||X_I|| = A_I\} \cong S_{A_1}^2 \times ... \times S^2_{A_4} $, is the space of intrinsic geometries of an open tetrahedron. Each $S^2_{A_I}$ is a 2-sphere with radius $A_I$, and $I \in \{1,2,3,4\}$. When the four vectors $X_I$ are constrained to sum to zero, the orthogonal surfaces associated to them close, giving a tetrahedron in Euclidean $\mathbb{R}^3$ with face areas $\{A_I\}$\footnote{Analogous arguments hold for the case of an open $d$-polyhedron and its associated closure condition.}. In this subsection we take $\Gamma_\ex = \Gamma_{\{A_I\}}$. \

The diagonal action of the Lie group $SU(2)$ (rotations) on $\Gamma_\ex$ has an associated momentum map $J : \Gamma_\ex \to \su(2)^*$ defined by,
\be \label{mom}
J = \sum_{I=1}^4 X_I 
\ee
where $||X_I|| = A_I$. Symplectically reducing $\Gamma_\ex$ with respect to $J=0$ level set gives the Kapovich-Millson phase space \cite{kapovich1996} $\mathcal{S}_4 = \Gamma_\ex//SU(2) = J^{-1}(0)/SU(2)$ of a closed tetrahedron with the given face areas, where notation $//$ means a symplectic reduction. It imposes closure of the four faces, with space $J^{-1}(0)$ being the constraint submanifold. But what we are interested in here is to define a Gibbs probability distribution on $\Gamma_\ex$ by imposing closure only on average, using the method of section \ref{maxent}. 

From a statistical perspective, the exact (or strong) fulfilment of closure can be understood as defining a microcanonical statistical state on $\Gamma_\ex$ with respect to this constraint. On the other hand, a weak fulfilment of the same constraint can be thought of as being implemented by a generalized Gibbs state. Their respective partition functions on the extended state space are then formally related by a Laplace transform.   

To define a Gibbs state with respect to closure for an open tetrahedron, we maximize the Shannon entropy \eqref{ent} under the constraints of state normalization and the following three,
\be \label{jconst}
\langle J_i \rangle_\rho \equiv \int_{\Gamma_\ex} d\lambda \; \rho \; J_i = 0 \hspace{0.5cm} (i=1,2,3) \;.
\ee
Here $\rho$ is a statistical state defined on $\Gamma_\ex$, and $J_i$ are components of $J$ in a basis of $\su(2)^*$. Clearly the above equation (for each $i$) is a weaker condition than imposing closure exactly by $J_i = 0$. Functions $J_i$ are smooth and real-valued on $\Gamma_\ex$, 
taking on the role of quantities $\mathcal{O}_a$ used in equations \eqref{const1}. Then optimizing the auxilliary functional of equation \eqref{auxfn} gives a Gibbs state on $\Gamma_\ex$ of the form,
\be \label{closuregibbs}
\rho_{\beta} = \frac{1}{Z{(\beta)}} e^{-\beta \cdot J} 
\ee
where $\beta \in \su(2)$ is a vector-valued temperature, with components $\beta_i$. Moreover, the function $\beta \cdot J= \sum_{i=1}^3 \beta_i J_i$ is the corresponding co-momentum map on $\Gamma_\ex$ (equivalently, the modular Hamiltonian). 

Evidently, the state $\rho_{\beta}$ is an example of a generalization of Souriau's Gibbs states \cite{souriau,e18100370}, to the case of Lie group actions of gauge symmetries generated by first class constraints, in a fully background independent setting. 




\subsection{Quantum statistical mechanics} \label{tet2}

In a quantum setting, each tetrahedron face $I$ is prescribed an $SU(2)$ representation label $j_I$ and Hilbert space $\mathcal{H}_{j_I}$. The tetrahedron itself is assigned an invariant tensor (intertwiner) of the four incident representation spaces. The full space of $4$-valent intertwiners is $\bigoplus_{j_I} \text{Inv} \otimes_{I = 1}^4 \mathcal{H}_{j_I}$, where Inv$\otimes_{I = 1}^4 \mathcal{H}_{j_I}$ is the space of $4$-valent intertwiners with given fixed spins $\{j_I\}$ (given fixed face areas), corresponding to a quantization of $\mathcal{S}_4$. A collection of neighbouring quantum tetrahedra has been associated to a 4-valent spin network \cite{Bianchi:2010gc}, with the labelled nodes and links of the latter being dual to labelled tetrahedra and their shared faces respectively of the former.  Then taking the viewpoint of tetrahedra as being extended `particles', the single particle Hilbert space of interest here is
\be
\mathcal{H} = \bigoplus_{j_I} \text{Inv} \otimes_{I = 1}^4 \mcH_{j_I} 
\ee
and, quantum states of a system of $N$ such tetrahedra are elements of $\mathcal{H}_N = \mathcal{H}^{\otimes N}$. We can equivalently work with the holonomy representation of the same quantum system in terms of $SU(2)$ group data, which is also the state space of a single gauge-invariant quantum of a group field theory defined on an $SU(2)^{4}$ base manifold \cite{Oriti:2006se,Oriti:2013aqa},
\be \label{singleH}
\mathcal{H} = L^2(SU(2)^{4}/SU(2)) \;.
\ee

Mechanical models of $N$ quantum tetrahedra can be defined by a set of gluing operators defined on $\mathcal{H}_N$. Thus a quantum mechanical model of a system of $N$ tetrahedra consists of the unconstrained Hilbert space $\mathcal{H}_N$, an operator algebra defined over it and a set of gluing operators specifying the model.

Now for a quantum multi-particle system, a Fock space is a suitable home for configurations with varying particle numbers. For bosonic (indistinguishable) quanta, each $N$-particle sector is the symmetric projection of the full $N$-particle Hilbert space, so that the Fock space is,
\be \mathcal{H}_F =  \bigoplus_{N \geq 0} \text{sym}\, \mathcal{H}_N \;.\ee
Fock vacuum $\ket{0}$ is the cyclic state with no tetrahedron degrees of freedom. Then, a system of an arbitrarily large number of quantum tetrahedra is described by the state space $\mathcal{H}_F$, an algebra of operators over it with a special subset of them identified as gluing constraints. Quantum statistical states of tetrahedra are density operators (self-adjoint, positive and trace-class) on $\mathcal{H}_F$ \cite{Kotecha:2018gof}. 

As before, a generalized Gibbs state in a Fock system of quantum tetrahedra with a constraint operator $\h{\mbC}$ is of the form,
\be\label{qeq} \h{\rho}_{\beta} = \frac{1}{Z_\beta} e^{- \beta \h{\mbC}} \ee 
where $\beta$ is associated with the condition $\langle \h{\mbC} \rangle = 0$. In particular, a density operator with a contribution from a grand-canonical weight of the form $e^{\mu \h{N}}$, will correspond to a situation with varying particle number, where $\h{N}$ is the number operator associated with the Fock vacuum. The corresponding partition function is 
\be\label{qgrand}
Z(\mu,\beta) = \Tr_{\mcH_F} \left[e^{- \beta\, \h{\mbC}\;\; + \;\; \mu \h{N} }\right]\;.
\ee
If $\h{\mbC}$ is a dynamical constraint of the system, which in general could include number- and graph-changing interactions, then one obtains a grand-canonical state of the type above with respect to $\h{\mbC}$.






\subsection{Field theory of quantum tetrahedra}

Hilbert space $\mcH_F$ is generated by ladder operators acting on the vacuum $\ket{0}$, and satisfying the algebra,
\be \label{ccr} [\h{\varphi}(\vec{g}),\h{\varphi}^*(\vec{g'})] = \delta(\vec{g},\vec{g'})\ee 
where $\delta$ is a delta distribution on the space of smooth, complex-valued $L^2$ functions on $SU(2)^4$, and $\vec{g} \equiv (g_1,...,g_4)$. 

For a state $e^{-\beta \h{C}}$, the traces in the partition function and other observable averages can be evaluated using an overcomplete basis of coherent states,
\be \ket{\psi} = e^{-\frac{||\psi||^2}{2}}e^{\int d\vec{g}\; \psi(\vec{g}) \h{\varphi}^*(\vec{g})}\ket{0} . \ee 
These states are labelled by $\psi \in \mcH$ and $||.||$ is the $L^2$ norm in the single particle Hilbert space $\mathcal{H}$. This gives,
\be \label{coherentZ}
\Tr({e^{-\beta\h{C}} \h{\mcO}}) = \int [D\mu(\psi,\bar{\psi})] \bra{\psi}e^{-\beta\h{C}}\h{\mcO}\ket{\psi} , \;\;\;\;\text{with}\;\; Z = \Tr(e^{-\beta\h{C}} \mathbb{I}) \;.
\ee 
Resolution of identity is $\mbI = \int [D\mu(\psi,\bar{\psi})] \ket{\psi}\bra{\psi}$, and the standard coherent state measure is 
$
D\mu(\psi,\bar{\psi}) = \prod_{k=1}^{\infty} d\,\text{Re}\psi_k \; d\,\text{Im}\psi_k / {\pi} \;.
$
The set of all such observable averages formally defines the total statistical system. 
In the following, we show how the quantum statistical partition function can be reinterpreted as the partition function for a field theory (of complex-valued $L^2$ fields) of the underlying quantum tetrahedra. \

For generic operators $\h{C}(\h{\varphi},\h{\varphi}^*)$ and $\h{\mcO}(\h{\varphi},\h{\varphi}^*)$ as polynomials in the algebra generators, and an arbitrary choice of operator ordering defining the exponential operator, the integrand of the statistical averages can be treated as follows.
\begin{align} \label{19}
\bra{\psi}e^{-\beta\h{C}}\h{\mcO}\ket{\psi} &= \bra{\psi}\sum_{k=0}^{\infty}\frac{(-\beta)^k}{k!}  \h{C}^k \h{\mcO}\ket{\psi}  \\
&= \bra{\psi} :e^{-\beta \h{C}} \h{\mcO}: \ket{\psi} + \bra{\psi} :\text{po}_{{C},{\mcO}}(\h{\varphi},\h{\varphi^*},\beta) : \ket{\psi}
\end{align}
where the second equality is gotten by using the commutation relations \eqref{ccr} on each $\h{C}^k\h{\mcO}$ and collecting all normal ordered terms $:\h{C}^k \h{\mcO}:$ to get the normal ordered $:e^{-\beta \h{C}} \h{\mcO}:$. The second term in the last line of \eqref{19} is a collection of the remaining terms arising as a result of exchanging $\h{\varphi}$'s and $\h{\varphi}^*$'s. In general, it will be a normal ordered series in powers of $\h{\varphi}$ and $\h{\varphi^*}$, with coefficient functions of $\beta$. The exact form of this series will depend on both $\h{C}$ and $\h{\mcO}$. 

Further recalling that coherent states are eigenstates of the annihilation operator, $\h{\varphi}(\vec{g})\ket{\psi} = \psi(\vec{g}) \ket{\psi}$, we have
\be
\bra{\psi} :e^{-\beta \h{C}}\h{\mcO}: \ket{\psi} = e^{-\beta C[\bar{\psi},\psi]} \mcO[\bar{\psi},\psi] 
\ee
where $C[\bar{\psi},\psi] = \bra{\psi}\h{C}\ket{\psi}$ and $\mcO[\bar{\psi},\psi] = \bra{\psi}\h{\mcO}\ket{\psi}$. Defining operators $\h{A}_{C,\mcO} \equiv \text{po}_{C,\mcO}(\h{\varphi},\h{\varphi^*},\beta)$, we have
\be
 \bra{\psi} :\h{A}_{C,\mcO}(\h{\varphi},\h{\varphi^*},\beta) : \ket{\psi} = A_{C,\mcO}[\bar{\psi},\psi,\beta]\; ,
\ee
encoding all higher order corrections. Averages \eqref{coherentZ} can thus be written as
\be \label{avgs} \Tr({e^{-\beta\h{C}} \h{\mcO}}) = \int [D\mu(\psi,\bar{\psi})] \; \left( e^{-\beta C[\bar{\psi},\psi]} \mcO[\bar{\psi},\psi] + A_{C,\mcO}[\bar{\psi},\psi,\beta] \right) \;. \ee
In particular, the quantum statistical partition function for a dynamical system of complex-valued $L^2$ fields $\psi$ defined on the base manifold $SU(2)^4$ is 
\be \label{pf} Z = \int [D\mu(\psi,\bar{\psi})] \; \left(e^{-\beta C[\bar{\psi},\psi]} + A_{C,\mbI}[\bar{\psi},\psi,\beta]\right) \;\equiv\; Z_{0} + Z_{\mcO(\hbar)} \;.
\ee
Here, by notation $\mcO(\hbar)$ we mean simply that this sector of the full theory includes all higher orders in quantum corrections relative to $Z_0$. This full set of observable averages (or correlation functions) \eqref{avgs}, including the above partition function, defines thus a statistical field theory of quantum tetrahedra (or in general, polyhedra with a fixed number of boundary faces), 
 characterized by a combinatorially non-local statistical weight. In other words, a group field theory. This statistical foundation of group field theories was first suggested in \cite{Oriti:2013aqa}.  
 
 Whenever it is possible to reformulate $A_{C,\mcO}$ such that
 \be A_{C,\mcO} = A_{C,\mbI}[\bar{\psi},\psi,\beta] \,\mcO[\bar{\psi},\psi]  \ee
 then \eqref{pf} defines a statistical field theory for the algebra of observables $\mcO[\bar{\psi},\psi]$. Further by rewriting $Z$ in terms of a simple exponential measure (under some approximations), we would get
\be
Z_{\text{eff}} = \int [D\mu(\psi,\bar{\psi})] \; e^{- C_{\text{eff}}[\bar{\psi},\psi, \beta, C, A]} \;,
\ee
making the correspondence with a standard field theory manifest.
A detailed discussion of the relation of the resulting statistical field theory to existing group field theory models, for topological BF theories, is given in \cite{Oriti:2013aqa,isha}.
 
\section{Conclusion}

We generalized Jaynes' information-theoretic approach of statistical equilibrium to a background independent system of many geometric tetrahedra. Using the symplectic description of classical tetrahedron geometry, we presented a natural generalization of Souriau's Gibbs states to a constrained system. Using a Fock space description, a quantum canonical partition function was put in relation with the generating function of labelled 2-complexes of discrete quantum gravity.

\end{document}